\documentclass[aps,prl,twocolumn,amsmath,amssymb,superscriptaddress]{revtex4}
\usepackage{graphicx}
\usepackage{amssymb}
\usepackage{natbib}
\usepackage{xfrac}
\usepackage{epstopdf}
\usepackage{float}
\usepackage{xspace}
\usepackage{xcolor}
\newcommand{\dxz}{$d_{xz}$\xspace}
\newcommand{\dyz}{$d_{yz}$\xspace}
\newcommand{\dxy}{$d_{xy}$\xspace}
\newcommand{\NFA}{NaFeAs\xspace}
\newcommand{\BFA}{BaFe$_2$As$_2$\xspace}

\begin{document}

\title{Orbital-selective spin waves in detwinned \NFA}

\author{David W. Tam}
\affiliation{Department of Physics and Astronomy, Rice University, Houston, Texas 77005, USA}
\author{Zhiping Yin}
\email{yinzhiping@bnu.edu.cn}
\affiliation{Center for Advanced Quantum Studies and Department of Physics, Beijing Normal University, Beijing 100875, China}
\author{Yaofeng Xie}
\affiliation{Department of Physics and Astronomy, Rice University, Houston, Texas 77005, USA}
\author{Weiyi Wang}
\affiliation{Department of Physics and Astronomy, Rice University, Houston, Texas 77005, USA}
\author{M. B. Stone}
\affiliation{Quantum Condensed Matter Division, Oak Ridge National Laboratory, Oak Ridge, Tennessee 37831, USA}
\author{D. T. Adroja}
\affiliation{ISIS Facility, Rutherford Appleton Laboratory, STFC, Chilton, Didcot OX11 0QX, United Kingdom}
\affiliation{Highly Correlated Matter Research Group, Physics Department, University of Johannesburg, P.O. Box 524, Auckland Park 2006, South Africa}
\author{H. C. Walker}
\affiliation{ISIS Facility, Rutherford Appleton Laboratory, STFC, Chilton, Didcot OX11 0QX, United Kingdom}
\author{Ming Yi}
\affiliation{Department of Physics and Astronomy, Rice University, Houston, Texas 77005, USA}
\author{Pengcheng Dai}
\email{pdai@rice.edu}
\affiliation{Department of Physics and Astronomy, Rice University, Houston, Texas 77005, USA}

\date{\today}

\begin{abstract}
The existence of orbital-dependent electronic correlations has been recognized as an essential 
ingredient to describe the physics of iron-based superconductors.
\NFA, a parent compound of iron based superconductors, exhibits
a tetragonal-to-orthorhombic lattice distortion below $T_s\approx 60$ K, forming
an electronic nematic phase with two 90$^\circ$ rotated (twinned) domains, and orders antiferromagnetically below $T_N\approx 42$ K.  We use inelastic neutron scattering to study spin waves  
in uniaxial pressure-detwinned \NFA.  By comparing the data with combined density functional theory and  
dynamical mean-field theory calculations, we conclude that spin waves up to an energy scale of 
$E_\text{crossover} \approx 100$ meV are dominated by
\dyz-\dyz intra-orbital scattering processes, which have the two-fold ($C_2$) rotational symmetry of the underlying lattice.
On the other hand, the spin wave excitations above $E_\text{crossover}$, which have approximately fourfold ($C_4$) rotational symmetry, arise from the \dxy-\dxy intra-orbital scattering that controls the overall magnetic bandwidth in this material. 
In addition, we find that the low-energy ($E\approx 6$ meV) spin excitations change from approximate $C_4$ to $C_2$ rotational symmetry below a temperature $T^\ast$ ($>T_s$), while spin excitations at energies above $E_\text{crossover}$ have approximate $C_4$ rotational symmetry and 
are weakly temperature dependent. These results are consistent with angle resolved photoemission spectroscopy measurements, 
where the presence of an uniaxial strain necessary to detwin \NFA also raises the onset temperature $T^\ast$ of observable orbital-dependent band splitting to above $T_s$, thus supporting the notion of orbital selective spin waves in the nematic phase of iron-based superconductors.  
\end{abstract}
\maketitle

% fig 1
\begin{figure*}[tbph]
\includegraphics[scale=.4]{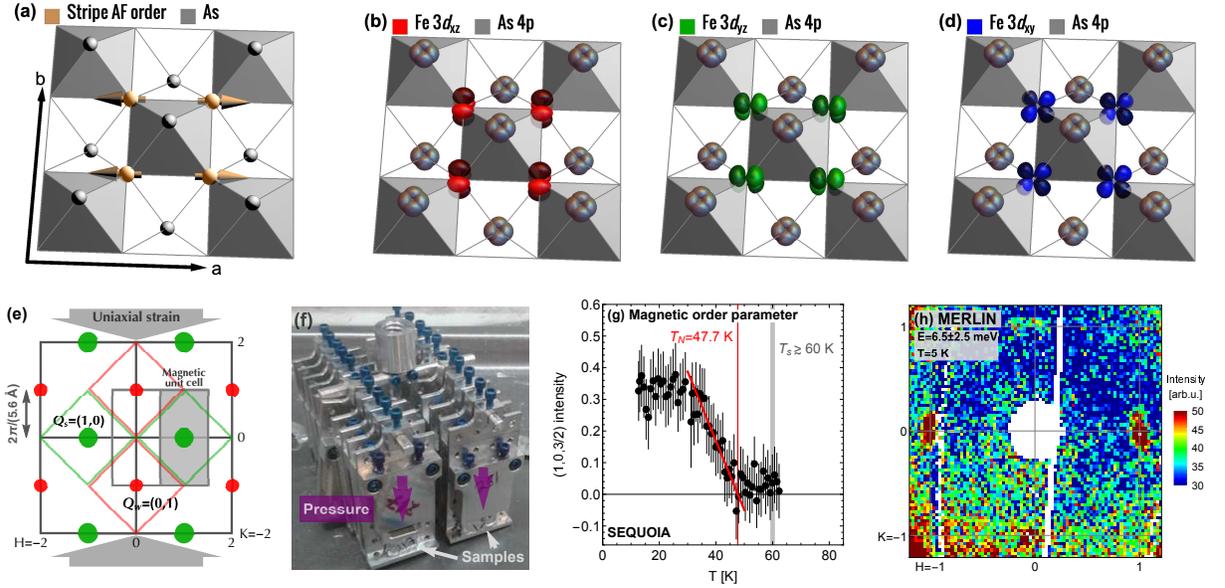}
\caption{
(a) Real-space spin structure of stripe-antiferromagnetically ordered \NFA.
(b-d) Fe $3d$ orbital wave functions projected into the real-space configuration, shown for \dxz in red, \dyz in green, and \dxy in blue, corresponding to the color scheme used throughout this work. The As $4p_x$ and $4p_y$ orbitals are shown in grey at each As site.
(e) In-plane ${\bf Q}$-space configuration in a detwinned sample with uniaxial compression along the vertical direction and the corresponding two-iron magnetic unit cell shaded in grey. Magnetic Bragg reflections at $\mathbf{Q}_\text{AF,strong}=(1,0)$ and equivalent are shown with green dots, corresponding to favorable nesting conditions for low-energy \dyz-\dyz intra-orbital scattering processes. Broken tetragonal symmetry below $T_s$ removes the equivalent nesting conditions for \dxz-\dxz processes, with the positions of residual magnetic Bragg peaks from the unfavored domains shown at $\mathbf{Q}_\text{AF,weak}=(0,1)$ and equivalent positions shown as smaller red dots. The diamond-shaped zones around each peak show the area of integrated intensity in Fig. 6.
(f) A picture of the 20 coaligned NaFeAs crystals in uniaxial pressure devices used in
the SEQUOIA and MERLIN experiments. The arrows 
mark the applied uniaxial pressure direction. (g) Temperature dependence of the elastic
scattering at the strong-domain magnetic Bragg peak position ${\bf Q}=(1,0,1.5)$. 
(h) A 2D image of $E=6.5\pm 2.5$ meV spin waves of NaFeAs at 10 K obtained on MERLIN
with incident neutron energy of $E_i=40$ meV along the $c$ axis.  
}
\end{figure*}

\begin{figure*}[tbph]
\includegraphics[scale=.4]{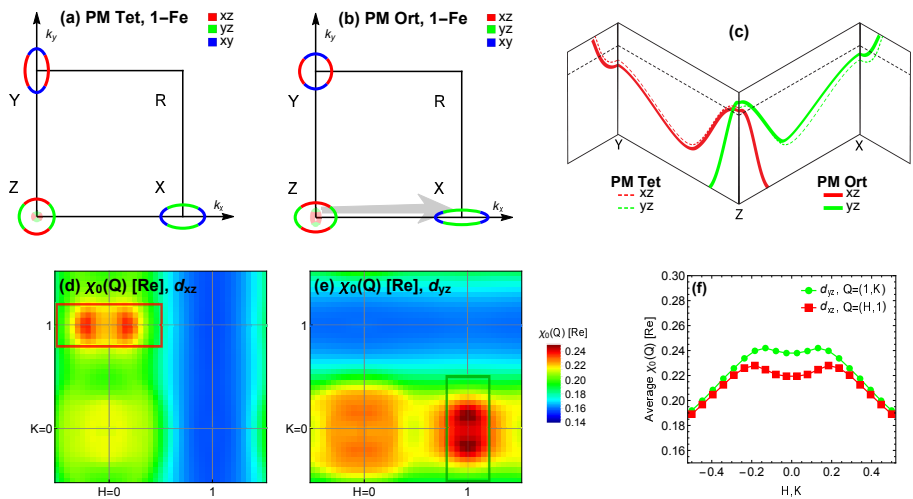}
\caption{
(a) Fermi surface nesting scheme in the high-temperature paramagnetic tetragonal phase and (b) detwinned paramagnetic orthorhombic phase stable for $T_N$ ($\approx 45$ K) $< T < T_s$ ($\approx 60$ K). The gray arrow in (b) shows the position of favorable nesting for \dyz-\dyz intra-orbital scattering at $\mathbf{Q}_\text{AF,strong}=(1,0,1)$ between the inner hole-like pocket at $Z=(0,0,1)$ and the elongated electron pocket at $X=(1,0,0)$. ARPES data indicates that the inner hole pocket is strictly closed at $Z=(0,0,1)$ in the high-temperature phase, and just touches the Fermi level in the paramagnetic orthorhombic phase. Therefore, we show this pocket in light shading to reflect the fact that nesting at $\mathbf{Q}_\text{AF,strong}=(1,0,1)$ may still be satisfied for very low energy scattering processes or for a slightly different choice of $k_z$ values, for example from $(0,0,0.9)$ to $(1,0,-0.1)$. (c) The effect of tetragonal-orthorhombic structure on 
the \dxz and \dyz orbital electronic 
structures of NaFeAs \cite{ZHYJ12,YLMK12}. 
(d-e) The calculated wavevector dependence of the real part 
of the bare susceptibility, $\chi_0(E,{\bf Q})$[Re], for (d) \dxz and (e) \dyz orbitals
at zero energy ($E=0$). (f) The intensity of $\chi_0(E,{\bf Q})$[Re] for 
\dxz and \dyz orbitals at $(0,1)$ and $(1,0)$ positions, respectively.   
}
\end{figure*}

% fig 2
\begin{figure*}[tbph]
\includegraphics[scale=.38]{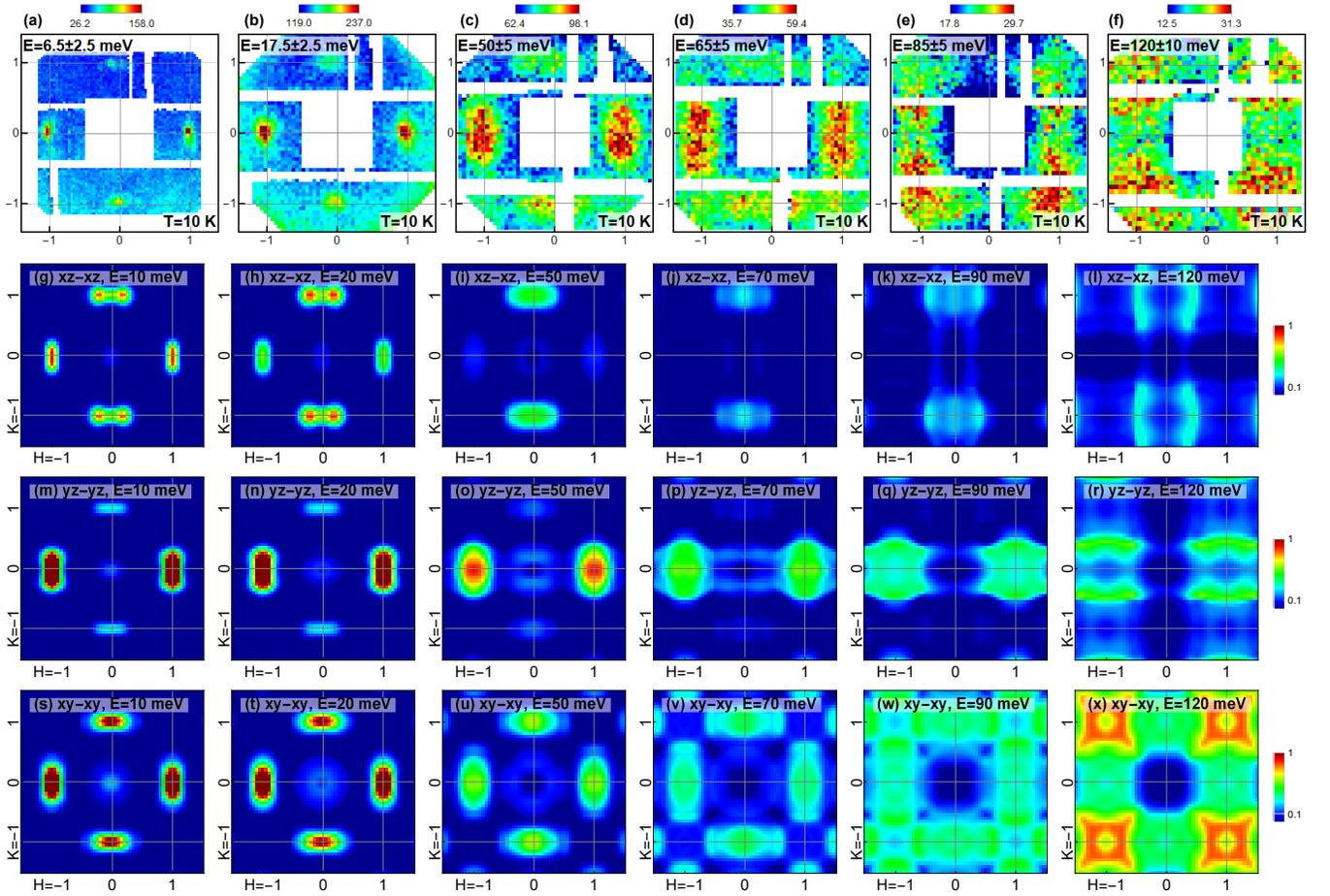}
\caption{ 
(a-f) Inelastic neutron scattering data from SEQUOIA at T=10 K, for energy transfer $E$ = $6.5 \pm 2.5$, $17.5 \pm 2.5$, $50 \pm 5$, $65 \pm 10$, $85 \pm 5$, and $120 \pm 10$ meV, in absolute units with color scale shown above.
Below, corresponding constant-energy slices from the DFT+DMFT calculation, in the \dxz-\dxz (g-l), \dyz-\dyz (m-r), and \dxy-\dxy (s-x) intra-orbital scattering channels, for similar energies compared with the neutron scattering data as shown in each panel, and the corresponding intensity scale to the right of each row.
The anisotropy of the \dyz-\dyz channel best resembles the low-energy measurements, but only \dxy-\dxy processes capture the migration of intensity to the ${\bf Q}=(1,1,L)$ position at high energies.
The \dxz-\dxz channel (second row) exhibits an orientation and anisotropy in the opposite sense from the data.
The color scale is fixed for all panels except the top row.
The $d_{xy}$ orbital data in (u-w) is broader in {\bf Q} and has large weakly dispersive background scattering, thus giving a comparable integrated intensity as that of the $d_{yz}$ orbital in the diamond-shaped zones depicted in Fig. 1(e) and Fig. 5(f), despite a smaller peak intensity at (1,0).
}
\label{fig2}
\end{figure*}

% fig 3
\begin{figure*}[tbph]
\includegraphics[scale=.48]{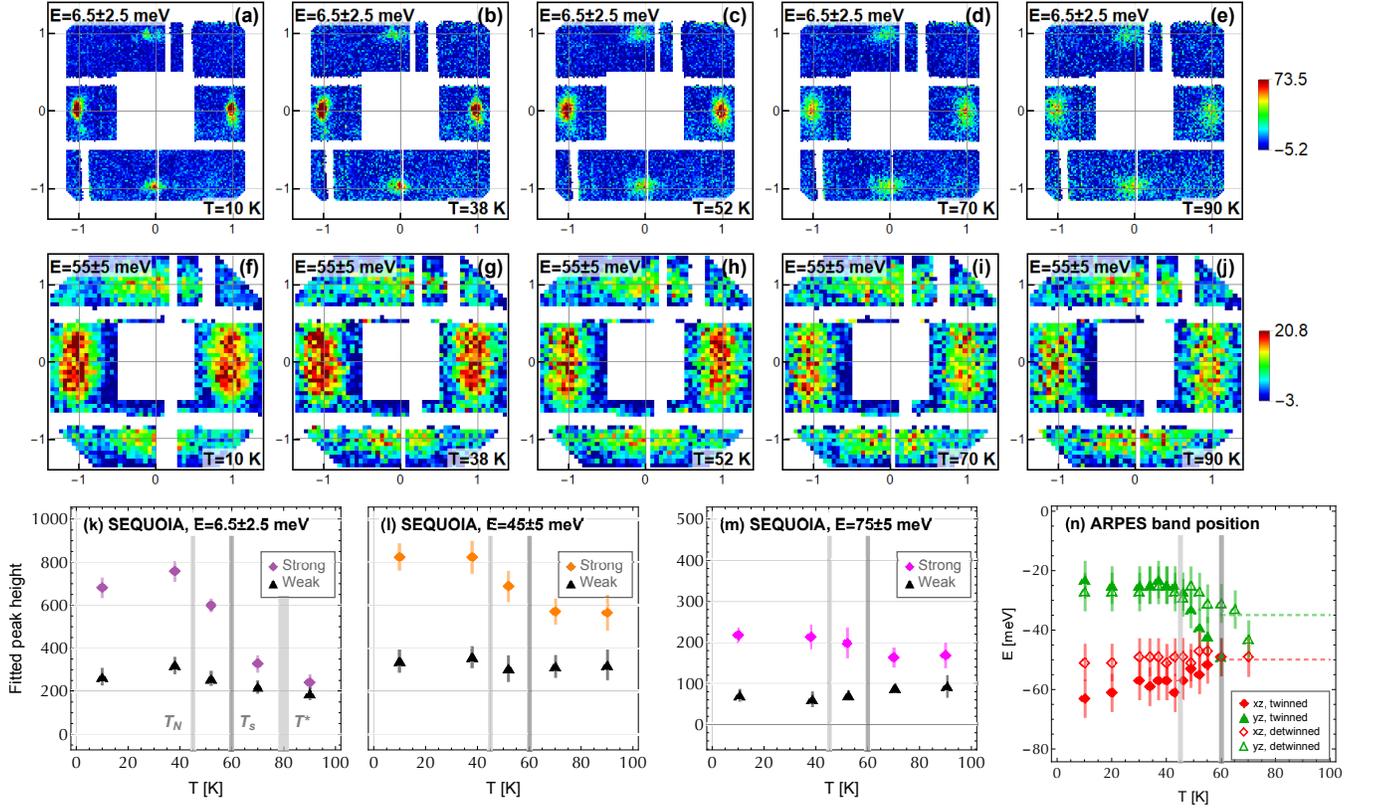}
\caption{
(a-e) Inelastic neutron scattering data from SEQUOIA at constant energy transfer $E$ = $6.5 \pm 2.5$ meV, at temperatures $T$=10, 38, 52, 70, and 90 K, after background subtraction and with constant color scale shown at right.
(f-j) The same as (a-e) at energy transfer $E$ = $55 \pm 5$ meV.
(k-l) Peak height of Gaussian fits to transverse cuts, at (k) $E=6.5 \pm 2.5$ meV and (l) $45 \pm 5$ meV, in absolute units.
(m) Peak height of the incommensurate (\dyz) peak, fitted along transverse cuts at $E=75 \pm 5$ meV to a two-peak model including one peak constrained to the position $\mathbf{Q}=(1,1)$. (n) Temperature dependence of the \dyz (green) and \dxz (red) band positions measured on twinned and detwinned crystals of NaFeAs from Ref. \cite{YLMK12}.
The light and dark shaded vertical lines in (k-n) mark zero pressure $T_N$ and $T_s$, respectively. The wide shaded line in (k) indicates $T^\ast$.} 
\end{figure*}

\begin{figure*}[tbph]
\includegraphics[scale=.48]{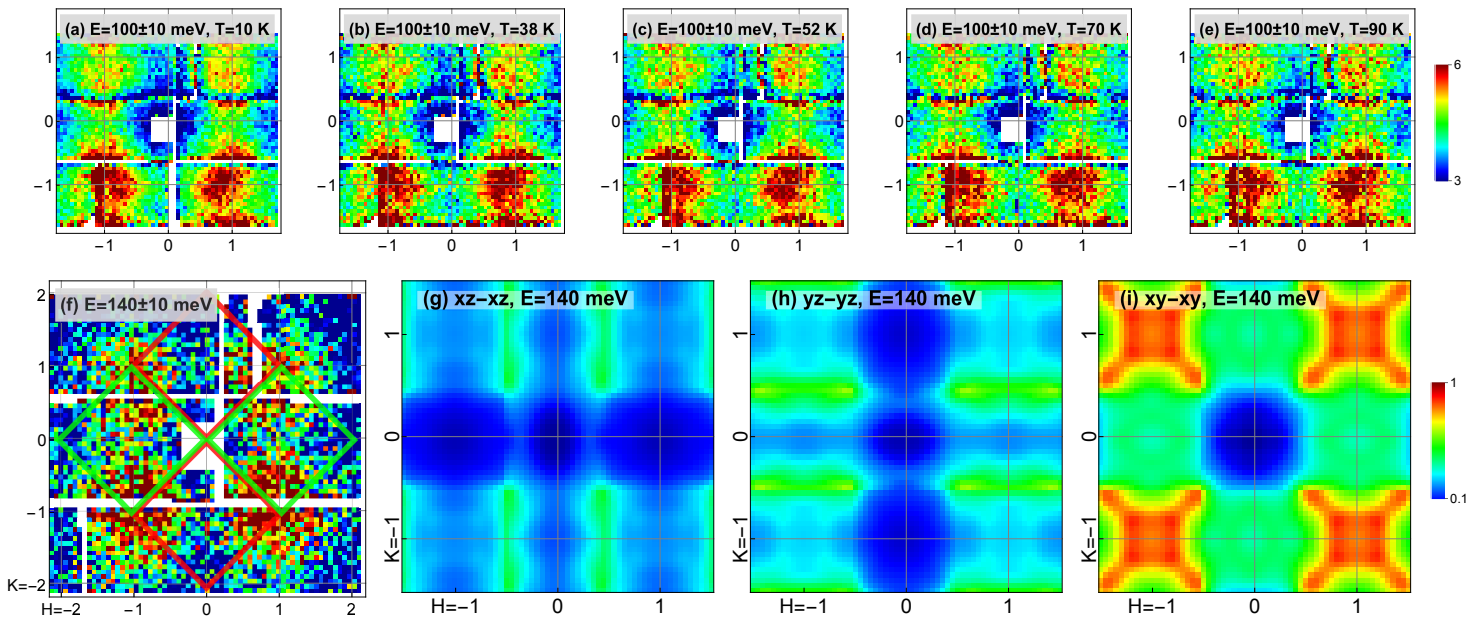}
\caption{(a-e) 2D images of spin excitations raw data of $E=100 \pm 10$ meV 
obtained on SEQUOIA with $E_i=150$ meV along the $c$ axis at  
temperatures $T=10, 38, 52, 70,$ and 90 K.
The streaks in the images represent the positions of detector gaps which are 
slightly {\bf Q}-dependent in the $E=90$-110 meV energy window at $E_i=150$ meV, thus leading to a smearing effect.
(f) 2D image of spin waves raw data of $E=140\pm 10$ meV at 10 K obtained on SEQUOIA
with $E_i=250$ meV along the $c$ axis. The green and red boxes are intensity integration regions for the strong and weak positions, respectively.
The DFT+DMFT calculation for $E=140\pm 10$ meV in the \dxz-\dxz (g), \dyz-\dyz (h), and \dxy-\dxy (i) intra-orbital scattering channels.
}
\end{figure*}

% fig 4
\begin{figure*}[tbph]
\includegraphics[scale=.55]{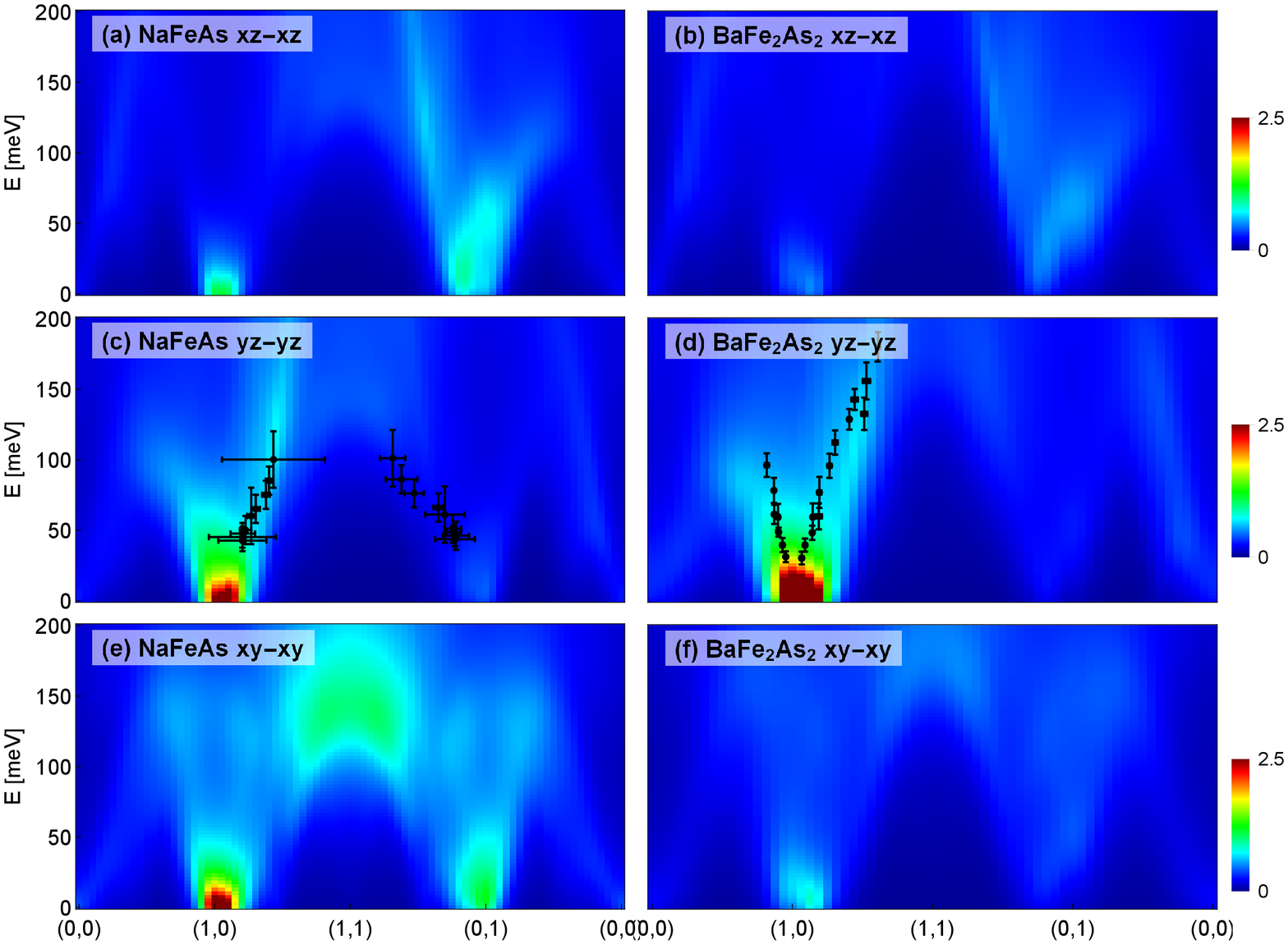}
\caption{
DFT+DMFT intensity along $\mathbf{Q}-E$ cuts around the high-symmetry paths in the in-plane orthorhombic unit cell, for \NFA (a,c,e) and \BFA (b,d,f), in (a-b) the \dxz-\dxz, (c-d) \dyz-\dyz, and (e-f) \dxy-\dxy intra-orbital scattering channels.
The peak positions from neutron scattering data closely match the DFT+DMFT intensity in the \dyz-\dyz channel for both (c) \NFA and (d) \BFA, with data taken from \cite{Lu18}.
Above $\sim$110 meV, spin fluctuations are observed at (1,1) and are consistent with the \dxy-\dxy scattering channel as shown in the constant-energy slices in Fig. \ref{fig2}.
Spin fluctuations at (1,1) are pushed above $E \approx 150$ meV in \BFA and persist to the band top of nearly 300 meV, showing that the larger overall bandwidth of \BFA is controlled by \dxy-\dxy fluctuations.
}
\end{figure*}

\begin{figure*}[tbph]
\includegraphics[scale=.55]{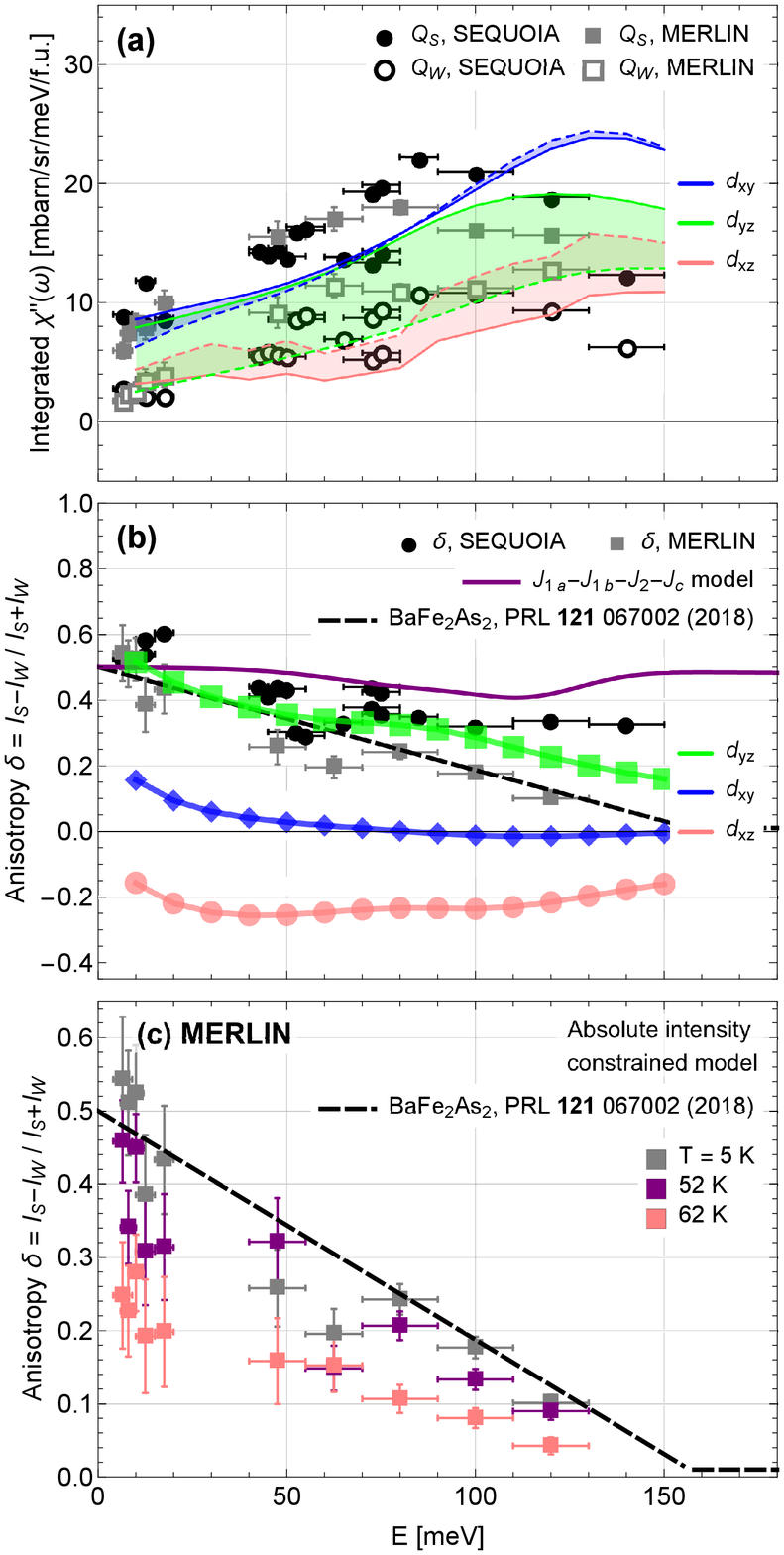}
\caption{ (a) Energy dependence of the momentum-integrated local dynamic magnetic susceptibility, from inelastic neutron scattering, for the strong (solid symbols) and weak positions (open symbols) at 10 K. Orbital selective DFT+DMFT calculation, with the strong-side integrated intensity shown as solid lines and weak-side intensity as dashed lines, demonstrates that \dyz-\dyz fluctuations (green) can explain the observed anisotropy.
(b) Anisotropy of the integrated intensity in part (b), demonstrating excellent agreement between experimental observations and \dyz-\dyz intra-orbital component of $\chi^{\prime\prime}$ computed by DFT+DMFT up to at least 100 meV. The purple line shows the results of linear spin wave theory using the Heisenberg parameters in \cite{Zhang14}.
(c) Anisotropy of spin excitations at 5 K ($\ll T_N$), 52 K ($>T_N$ and $<T_s$), 
and 62 K (larger than zero pressure $T_s$).  The dashed black lines in (b,c) 
are the results obtained for \BFA reported in \cite{Lu18}.
}
\end{figure*}

\section{introduction} 
High temperature superconductivity in the copper oxides and iron pnictides arises from electron- and hole-doping of their respective antiferromagnetic (AF) ordered parent compounds \cite{scalapinormp,dairmp}.  In the case of copper oxide
superconductors, the parent compounds are Mott insulators with strong electron
correlations \cite{palee}, and superconductivity arises from 
 electron pairing in the CuO$_2$ plane from the one-electron $d_{x^2-y^2}$ Cu orbitals \cite{scalapino}. 
For iron pnictide superconductors, the parent compounds such as BaFe$_2$As$_2$ and NaFeAs 
are semi-metallic \cite{stewart}, and antiferromagnetism can be generated either by Fermi surface 
nesting of itinerant electrons \cite{hirschfeld}, or local moments mediated by superexchange interactions through the As anions [Fig. 1(a)] \cite{QSi}.  Instead of the single $d_{x^2-y^2}$ band 
in copper oxides, the electronic structure of iron pnictides near the Fermi level contains several active bands with appreciable mixing of the 
Fe 3$d$ $t_{2g}$ orbitals (\dxz, \dyz, and \dxy) [Figs. 1(b-d)] \cite{YZSL17}. Therefore, 
a fundamental difference between copper oxides and iron pnictides is the multi-orbital multi-band nature of the underlying electronic structure in iron-based superconductors.

In addition to having collinear AF order below $T_N$ as shown in Fig. 1(a), 
iron pnictides exhibit
a tetragonal 
($C_4$ symmetric) to orthorhombic ($C_2$ symmetric) lattice 
distortion at temperature below $T_s$ ($T_s\geq T_N$) \cite{stewart}, forming an electronic nematic phase with two 90$^\circ$ rotated (twinned) domains [Figs. 1(e-h)] \cite{Yin11,Fernandes}. In the paramagnetic tetragonal ($C_4$) and nematic orthorhombic ($C_2$) phases, the Fermi surfaces of \NFA observed by angle resolved photo-emission spectroscopy (ARPES) experiments are schematically shown in 
Figs. 2(a) and 2(b), respectively \cite{YZSL17}. The appearance of a nematic phase 
 is associated with 
an energy splitting between \dxz and \dyz orbitals, where 
the \dyz band of the electron Fermi
surface at $X$ goes up in energy, while the \dxz band at $Y$ goes
down in energy [Fig. 2(c)] \cite{YZSL17}.
The splitting leads to more favorable Fermi surface nesting conditions for the \dyz orbital along the AF 
ordering direction [Figs. 1(e) and 2(b)], and lowers the rotational symmetry of electronic interactions that give rise to anisotropic spin fluctuations \cite{WKRE16}; conversely, spin fluctuations with lower rotational symmetry may themselves generate rotational anisotropy in the orbital channel \cite{FMTC16}.
The situation is not necessarily the same for strongly-coupled, localized electrons, through which spin fluctuations are generated by superexchange interactions taking place between Fe ions across the anion (pnictogen/chalcogen) site, reminiscent of strongly-coupled electrons in the cuprate materials \cite{QSi}.
In the iron-based superconductors, the \dxy orbital is most closely positioned to the anions, and it is also the most strongly localized \cite{YZSL17}.
For example, in FeTe$_{1-x}$Se$_x$ superconductors with a nematic phase 
but without static AF order \cite{Hsu,McQueen09,bohmer}, signatures of the \dxy orbital completely vanish from ARPES measurements at high temperature, indicating that the \dxy orbital is becoming more localized, possibly entering into an orbital-selective Mott phase 
 \cite{QSi,YLZY15} or a highly incoherent state through the formation of large on-site fluctuating moments due to the Hund's rule coupling \cite{HMiao2016,Yin2012}.
From scanning tunneling experiments on FeSe, superconductivity is believed to occur via orbital-selective Cooper
pairing from the \dyz orbital submanifold \cite{Sprau}.

In the case of \NFA, ARPES studies have found that the \dxy-dominated band is more strongly renormalized than the same band in other pnictides, but less so than chalcogenides.
Thus, one has a system with moderate overall correlation strength \cite{Yin11,YZSL17} that may exhibit spin fluctuations from two distinct mechanisms with different intra-orbital scattering 
processes of different correlation strength.
Differentiation of these mechanisms has important consequences for the origins of 
superconductivity \cite{ChKF16}.
Inelastic neutron scattering experiments on detwinned 
FeSe \cite{TChen2019}, underdoped superconducting 
NaFe$_{0.985}$Co$_{0.015}$As \cite{WYwang2017}, and Ba(Fe$_{1-x}$Co$_x$)$_2$As$_2$ \cite{Tian19}
suggest that the neutron spin resonance, a collective magnetic excitation coupled  
to the superconducting order parameter \cite{Eschrig,Maier,RZhang,Inosov2010,CHLee2013,Fwaber2019}, likely arises from orbital-selective electron-hole Fermi surface nesting of quasiparticles within the single \dyz orbital submanifold.

In previous inelastic neutron scattering work on twinned NaFeAs \cite{Zhang14} 
and $A$Fe$_2$As$_2$ ($A=$ Ca, Ba, Sr) \cite{JZhao09,leland11,Ewings11}, 
it was found that spin waves in NaFeAs have a zone boundary energy lower than 
that of $A$Fe$_2$As$_2$.  These results have been interpreted as due to 
increased electron-electron correlations from larger pnictogen height \cite{Yin11}, 
consistent with a combined density functional theory (DFT) 
and dynamical mean field theory (DMFT) calculation \cite{Zhang14,Yin14}. As these results are obtained
on twinned samples that 
obey the $C_4$ rotational symmetry,
one cannot deduce the relationship between spin excitations and orbital degrees of freedom from symmetry considerations alone in the twinned case.
Although recent spin wave 
measurements on detwinned BaFe$_2$As$_2$ reveal strong magnetic anisotropy between the AF 
wave vector $\mathbf{Q}_\text{strong}=(1,0)$ and $\mathbf{Q}_\text{weak}=(0,1)$ in 
the collinear AF state below the ordering temperature 
$T_N$ \cite{Lu18}, the nearly coupled structural 
and AF phase transitions in BaFe$_2$As$_2$ \cite{qhuang,kim2011} prevent 
a determination of the relationship between the $\mathbf{Q}_\text{strong}=(1,0)$ and $\mathbf{Q}_\text{weak}=(0,1)$ 
magnetic anisotropy and nematic (structural) phase transition below $T_s$.
On the other hand, \NFA exhibits a tetragonal-to-orthorhombic 
structural transition at $T_s\approx 60$ K and 
a separate collinear AF order below $T_N\approx 42$ K at ${\bf Q}_{AF}=(1,0)$ 
[Fig. 1(g)] \cite{SLi09,Tan16}.
Thus, detwinned \NFA not only presents a platform for testing the relationship between itinerant and localized physics in different orbital degrees of freedom, but also tests how spin waves are associated with the nematic phase across $T_s$.

\section{Experimental Details}

Our inelastic neutron scattering experiments are carried out at the SEQUOIA \cite{SQUOIA} and MERLIN \cite{MERLIN} neutron time-of-flight chopper spectrometers, located, respectively, at the Spallation Neutron Source, Oak Ridge National Laboratory, and ISIS facility, Rutherford-Appleton Laboratory. 
We define the wave vector \textbf{Q} in three-dimensional reciprocal space in \AA$^{-1}$ as ${\bf Q}=H {\bf a^\ast}+K{\bf b^\ast}+L{\bf c^\ast}$, where $H$, $K$, and $L$ are Miller indices and ${\bf a^\ast}=\hat{{\bf a}}2\pi/a, {\bf b^\ast}=\hat{{\bf b}} 2\pi/b, {\bf c^\ast}=\hat{{\bf c}}2\pi/c$ are reciprocal lattice units (r.l.u.) of the orthorhombic unit cell, where $a\approx 5.6$ \AA, $b\approx 5.6$ \AA, and $c\approx 7.0$ {\AA} \cite{SLi09,Tan16,SI}.  
Figures 1(a) and 1(e) show schematic diagram of the collinear AF order in NaFeAs in real 
and reciprocal space, respectively. In the unstrained state, \NFA forms 
twin structural domains below $T_s$ [Fig. 1(e)]. 
To overcome this, uniaxial mechanical pressure has been used to favor 
one twin orientation \cite{Dhital2012,Dhital2014,YSong2013,Tam2017,Lu14}.
Twenty single crystals of \NFA were aligned and cut into rectangular shapes along the $a/b$ direction, individually potted in hydrogen-free Cytop M glue and wrapped in aluminum foil to protect them from exposure to air [Fig. 1(f)]. 
Since the NaFeAs single crystals are detwinned via  
uniaxial pressure along the $b$-axis [the vertical arrow direction 
using aluminum springs in an aluminum manifold as shown in Fig. 1(f)], the AF order occurs at the in-plane wave vector 
$\mathbf{Q}_\text{strong}=(\pm 1,0)$ positions, and 
there should be no magnetic scattering at $\mathbf{Q}_\text{weak}=(0,\pm1)$ positions. For a 100\% detwinned 
sample, the magnetic Bragg peak 
intensity at the AF 
in-plane wave vector $\mathbf{Q}_\text{strong}$  
should be a factor of 2 of that 
at zero pressure, while there should be no observable magnetic Bragg scattering
at the $\mathbf{Q}_\text{weak}$ position [Fig. 1(e)]. 
In typical neutron triple-axis experiments on one single crystal sample of iron pnictides, the detwinning ratio of the system can be accurately determined by either comparing the uniaxial pressured induced magnetic Bragg peak intensity gain at 
$(\pm 1,0,L)$ or comparing the magnetic Bragg peak intensities at $(\pm 1,0,L)$ and $(0,\pm1,L)$ positions, where $L=1, 3, \cdots$ for BaFe$_2$As$_2$ and $0.5, 1.5,\cdots$ for NaFeAs 
\cite{Dhital2012,Dhital2014,YSong2013,Tam2017}.  Unfortunately, the rectangular assembly of 20 single crystals of NaFeAs under uniaxial pressure shown in Fig. 1(f) could not be arranged in the sample environment of 
a triple-axis experiment.  For our measurements using neutron time-of-flight spectrometers with only one sample rotation axis ($A3$, along the $b$ axis), elastic scans at different temperatures were collected by rotating the $A3$ angle of the sample assembly to satisfy the elastic condition ($E=0$) at the magnetic $(1,0,1.5)$ Bragg reflection and  
revealed $T_N\approx 47$ K [Fig. 1(g)] \cite{SLi09}. Since it is not possible to rotate the ``weak'' magnetic Bragg peak at $(0,\pm1,1.5)$ positions onto the detector bank at fixed $E_i$, one cannot use the above mentioned method to directly measure the sample assembly detwinning ratio, defined as 
$\eta=[I(1,0,1.5)-I(0,1,1.5)]/[I(1,0,1.5)+I(0,1,1.5)]$, where 
$I(1,0,1.5)$ and $I(0,1,1.5)$ correspond to the strong and weak magnetic Bragg peaks, respectively \cite{WYwang2017}.
 Nevertheless, one can still determine the sample detwinning ratio $\eta$ using the 
low-energy inelastic part of the spin fluctuation spectrum, which at least provides a firm lower bound, and was in fact observed to have one-to-one correspondence to $\eta$ in previous experiments on the NaFeAs and BaFe$_2$As$_2$ systems \cite{WYwang2017,Tian19}.  

We estimate the detwinning ratio $\eta$ by integrating the inelastic fluctuation spectrum for energies just above the spin anisotropy gap where spin waves appear at both the strong and weak positions.
For spin wave energies below 20 meV, the anisotropy is not highly energy-dependent, giving us confidence in this method.
From the inelastic slice containing $4 \leq E \leq 9$ meV 
with $E_i=40$ meV [Fig. 1(h)], we find that the integrated intensities at 
the strong ($I_s$ at $\mathbf{Q}_\text{strong}$) and weak ($I_w$ at $\mathbf{Q}_\text{weak}$) positions can give $\eta\approx 
(I_s-I_w)/(I_s+I_w)=50\%$, or a 3:1 domain ratio, only slightly
smaller than $\eta\approx 62.4\%$ obtained on one single crystal of NaFe$_{0.985}$Co$_{0.015}$As \cite{WYwang2017}.

Our measurements were carried out with incident neutron energies $E_i=$40, 80, 150, and 250 meV at SEQUOIA, and $E_i=$29, 54, and 170 meV at MERLIN, with the incident beam always along the $c$ axis.
In this experimental geometry, points with a given $H$, $K$, and energy transfer are always coupled to specific values of $L$. A typical constant energy slice contains data from a wide range of $L$ values spanning at least half of a unit cell along the $c$ axis direction. We simply average over the $L$ values contained in a given slice, converting the analysis into a two-dimensional (2D) problem.
For simplicity, we refer to the $(H,K)$ values without specifying $L$ when the energy and $E_i$ are also specified.

\section{Experimental Results}

Figures 2(a) and 2(b) schematically illustrate the electron-hole Fermi surfaces of NaFeAs and their nesting conditions above and below $T_s$ for different orbitals 
 \cite{WKRE16,ZYLL16,ZHYJ12,YLMK12}. In the paramagnetic tetragonal 
state, the electron-hole Fermi surface
nesting of the \dyz-\dyz and \dxz-\dxz orbital quasiparticles is near
the $\mathbf{Q}_\text{strong}=(\pm 1,0)$ and $\mathbf{Q}_\text{weak}=(0,\pm1)$ positions, respectively. 
When the system is cooled below $T_s$ in the paramagnetic orthorhombic nematic phase, 
the \dyz electron band shifts up in energy while the \dxz electron band shifts down [Fig. 2(c)] \cite{ZHYJ12,YLMK12}. 
As a result, the inner hole-like pocket at the zone center that is partially closed along $k_z$ acquires an elliptical shape which improves the nesting condition close to the Fermi level due to the larger \dyz spectral weight near $E=0$, therefore enabling scattering processes involving the \dyz orbitals to participate in fluctuations with lower overall energy at approximately $\mathbf{Q}_\text{strong}=(\pm 1,0)$.
Similarly, scattering processes with wavevector close to $\mathbf{Q}_\text{weak}$ direction are more favorable for \dxz-\dxz orbital 
hole-electron Fermi surface scattering \cite{JHZhang2010}.

To determine quantitatively if the above Fermi-surface nesting description is appropriate for understanding the collinear AF order in NaFeAs, we calculated the 
wavevector ($\bf Q$) dependence of 
the real part of the bare susceptibility, $\chi_0(E,{\bf Q})$[Re] \cite{Johannes06}, 
at zero energy ($E=0$) contributed by the Fe \dxz and \dyz orbitals (intra-orbital components). While the Fe \dxz orbital component has a peak structure around $\mathbf{Q}_\text{weak}$ [Fig. 2(d)], the Fe \dyz orbital component peaks around 
$\mathbf{Q}_\text{strong}$ [Fig. 2(e)]. Since the comparison plot in Fig. 2(f) 
shows a somewhat larger ($\approx 10$ \%) bare susceptibility around $(1, 0)$ than that around $(0,1)$, the Fermi surface nesting favors
$\mathbf{Q}_\text{strong}$ in this compound. However, 
the Fermi surface nesting itself cannot produce the large anisotropy of the experimental spin excitations between $(1, 0)$ and $(0, 1)$ [Fig. 1(h)]. 
 The two-particle vertex correction strongly enhances this difference and gives rise to the large anisotropy of the spin susceptibility between $(1, 0)$ and $(0, 1)$, defined as $\chi(E,{\bf Q})=\chi_0(E,{\bf Q})/[1-\Gamma \chi_0(E,{\bf Q})]$ where $\Gamma$ is the two-particle vertex function \cite{Yin14}. For example, if 
$\Gamma \chi_0(E,{\bf Q})=0.99$ at $(1, 0)$ and $\Gamma \chi_0(E,{\bf Q})=0.90$ at $(0, 1)$, the spin susceptibility $\chi(E,{\bf Q})$ at $(1,0)$ is about 10 times the value at $(0, 1)$. Therefore, both the Fermi surface nesting and the two-particle vertex correction play important roles in forming the AF order at $\mathbf{Q}_\text{strong}$.

Figures 3(a-f) show 2D images of spin waves 
at constant energy transfers of $E=6.5$, 17.5, 50, 65, 85, and 120 meV in the AF ordered state at 10 K. 
At $E=6.5\pm 2.5$ meV, there is clear spin wave anisotropy at 
$\mathbf{Q}_\text{strong}=(\pm 1,0)$ and 
$\mathbf{Q}_\text{weak}=(0,\pm1)$ positions.  
With increasing energy transfer, spin wave anisotropy between $\mathbf{Q}_\text{strong}$ 
and $\mathbf{Q}_\text{weak}$ becomes smaller and spin waves become almost $C_4$ rotational symmetric
around $E=120\pm 10$ meV. In addition, spin waves broaden and split in the transverse direction with increasing energy, and merge into the $\mathbf{Q}=(1,1)$ position where they remain somewhat anisotropic at $E=120\pm 10$ meV.  
Figures 3(g-l), 3(m-r), and 3(s-x) show the corresponding cuts in the 
 \dxz-\dxz, \dyz-\dyz, and \dxy-\dxy intra-orbital scattering channels, respectively, from the combined DFT+DMFT calculations. 
For spin waves from the \dxz-\dxz orbital quasiparticle excitations [Figs. 3(g-l)], the spectral weight of
the spin waves are mostly focused at $\mathbf{Q}_\text{weak}$, inconsistent with the observed low-energy
spin waves [Figs. 3(a-d)]. On the other hand, spin waves from  the \dyz-\dyz orbital 
quasiparticle excitations are opposite and 
have large spectral weight at $\mathbf{Q}_\text{strong}$ positions [Fig. 3(m-r)].  
Finally, spin waves from the \dxy-\dxy orbital
scattering have spectral weight at both $\mathbf{Q}_\text{strong}$ and $\mathbf{Q}_\text{weak}$ with 
approximate $C_4$ rotational symmetry [Fig. 3(s-x)].

In previous inelastic neutron scattering work 
on twinned NaFeAs, low energy spin excitations are
enhanced dramatically below $T_s$ instead of $T_N$ due to spin-lattice coupling \cite{LYSW18}.
To determine if spin excitation anisotropy at $\mathbf{Q}_\text{strong}$ and $\mathbf{Q}_\text{weak}$ in 
NaFeAs is also controlled by the nematic phase below $T_s$,  
we summarize in Fig. 4 temperature dependence of the spin excitations at different energies across $T_N$ and $T_s$.
Figures 4(a-e) and 4(f-j) plot temperature evolution of spin waves at $E=6.5\pm 2.5$ meV and $E=55\pm 5$ meV, respectively. 
Figures 4(k), 4(l), and 4(m) show temperature dependence of the spin excitation peak intensity at  
$\mathbf{Q}_\text{strong}$ and $\mathbf{Q}_\text{weak}$ for energies $E=6.5\pm 2.5$, $E=45\pm 5$, $E=75\pm 5$ meV, respectively. Since the widths of spin excitations at these energies are identical for $\mathbf{Q}_\text{strong}$ and $\mathbf{Q}_\text{weak}$ and only weakly 
temperature dependent as shown in the supplementary information \cite{SI}, temperature dependence of the peak intensity should give an accurate account of the spin excitation anisotropy. 

For spin excitations at $E=6.5\pm 2.5$ meV, we find enhanced magnetic scattering 
at a temperature $T^\ast$ above $T_s$ marked as a wide dark line, different from that of previous work on unstrained twinned NaFeAs \cite{LYSW18}.  
 In addition, a strong 
magnetic excitation anisotropy between $\mathbf{Q}_\text{strong}$ and $\mathbf{Q}_\text{weak}$ is also established at temperatures across $T_s$, and persists to temperatures above $T_s$ [Fig. 4(k)]. 
Upon increasing excitation energies to $E=45\pm5$ [Fig. 4(l)]
and $75\pm5$ [Fig. 4(m)] meV, spin excitation anisotropy is also present 
to temperatures above $T_s$.
These results are consistent with ARPES  
and transport measurements on both NaFeAs and Ba(Fe$_{1-x}$Co$_x$)$_2$As$_2$,
where anisotropy reflecting the nematic order is observed to 
on-set at a temperature $T^\ast$ above the zero pressure $T_s$ due to the
applied uniaxial pressure necessary to detwin and 
maintain the single domain orthorhombic AF phase of iron
pnictides \cite{ZHYJ12,YLMK12,Chu10,ZYLL16,HRMan18,DWTam2019}.
 To see this, Figure 4(n) compares the temperature dependence of the band splittings in pressure-free and uniaxial pressure-strained NaFeAs single crystals \cite{YLMK12}. 
Without applied uniaxial pressure, the band splitting of an unstrained (twinned) NaFeAs crystal onsets at $T_s$. 
For a detwinned crystal in the presence of uniaxial strain, the band 
splitting due to orbital anisotropy is resolvable up to at least 70 K, well above $T_s$ \cite{YLMK12} [Fig. 4(n)]. 
 The comparison with and without strain shows that this elevation of band splitting to higher temperatures is a direct result of uniaxial strain, which acts like a biasing field to the nematic order just like a magnetic field does to a ferromagnetic order. It is likely that a small anisotropy persists as long as the biasing field---strain in this case---is present, but unresolvable due to finite energy resolution in the ARPES experiments. This is consistent with the observation 
that while low energy spin excitations in BaFe$_2$As$_2$ 
($T_N\approx T_s\approx 140$ K)  have a small anisotropy 
at all temperatures below 250 K, the enhancement of 
spin excitation anisotropy below $T^\ast$ ($T_N, T_s<T^\ast \ll 250$ K) is associated with resistivity 
anisotropy and band splitting \cite{DWTam2019}.
Although the exact uniaxial pressures used in Ref. \cite{YLMK12} and Figs. 4(k-m) are unknown, it is clear that the $T^\ast$ associated with the enhanced spin excitation anisotropy occurs at approximately the band splitting temperature of a strained NaFeAs.

To test temperature dependence of the 
spin excitation anisotropy for energies above $E=80$ meV, 
we show in Fig. 5(a-e) 2D raw images of the $E=100\pm 10$ meV spin excitations at
$T=10, 38, 52, 70, 90$ K, respectively, obtained on SEQUOIA with $E_i=150$ meV. On warming 
across $T_N$ and $T_s$, spin excitations  
have approximate $C_4$ rotational symmetry with weak anisotropy, but are essentially temperature independent. Figure 5(f) shows the 2D raw image of spin excitations
at $E=140\pm 10$ meV obtained 
on SEQUOIA with $E_i=250$ meV and $T=10$ K, again revealing  
approximate $C_4$ rotational symmetry with weak anisotropy. 
Comparing these results with the combined DFT+DMFT calculations at $E=140$ meV
for \dxz-\dxz [Fig. 5(g)], \dyz-\dyz [Fig. 5(h)], and \dxy-\dxy [Fig. 5(i)] intra-orbital scattering channels, we see that spin excitations 
with energies above $E=100$ meV are peaked at the $(\pm 1,\pm 1)$ positions,
which is most consistent with the \dxy-\dxy intra-orbital scattering [Fig. 5(i)].

To understand the small, weakly temperature dependent, 
spin excitation anisotropy observed at all energies in Figs. 4 and 5, we note that 
the applied uniaxial pressure necessary to detwin the sample also induces an orthorhombic lattice distortion and a strain field at 
all temperatures \cite{XYLu2016}. For an applied pressure of
$P\approx 20$ MPa, the pressure-induced lattice parameter distortion 
in BaFe$_2$As$_2$ at temperatures well above 
the zero pressure $T_s$ is weakly temperature dependent and about
15\% of the intrinsic orthorhombic lattice distortion below $T_s$ 
[see Fig. 1(c) of Ref. \cite{XYLu2016}]. Since the resistivity and spin excitation anisotropy
at temperatures above the zero pressure $T_s$ increase with increasing 
uniaxial pressure \cite{Chu10,HRMan18,DWTam2019}, 
it is safe to assume that the uniaxial 
pressure-induced strain field in NaFeAs will produce a small $C_4$ rotatinal symmetry-breaking spin excitation component at all energies and temperatures.

\section{DFT+DMFT calculations and comparison with data}

DFT+DMFT \cite{Kotliar2006} was employed to compute the electronic structure and spin dynamics of 
NaFeAs and BaFe$_2$As$_2$ in the nematic state. The full-potential linear augmented plane wave method implemented in Wien2K \cite{Blaha2001} was used for the density functional theory part in conjunction with Perdew-Burke-Ernzerhof generalized gradient approximation \cite{Perdew1996} of the exchange correlation functional. DFT+DMFT was implemented on top of Wien2K and documented in Ref. \cite{Haule2010}. In the DFT+DMFT calculations, the electronic charge was computed self-consistently on DFT+DMFT density matrix. The quantum impurity problem was solved by the continuous time quantum Monte Carlo method \cite{Haule2007,Werner2006} with a Hubbard $U=5.0$ eV and Hund's rule coupling $J_H=0.8$ eV in the paramagnetic state \cite{Yin11,Yin14}.  Such a choice of the Hund’s rule coupling is essential for the large spin excitation strength and small spin-wave bandwidth. If the Hund’s rule coupling is small, there would be very small local spin moment, very weak spin excitations, and large spin-wave bandwidth. There are also hybridizations of the \dxz, \dyz, and \dxy orbitals between the neighboring Fe atoms, and therefore to the inter-orbital contributions of the total spin excitations. Since the strength of the inter-orbital contribution lies between the individual intra-orbital contributions, 
we can focus on the intra-orbital contribution to obtain a simplified but largely accurate picture.

The Bethe-Salpeter equation is used to compute the dynamic spin susceptibility where the bare susceptibility is computed using the converged DFT+DMFT Green’s function while the two-particle vertex is directly sampled using continuous time quantum Monte Carlo method after achieving full self-consistency of DFT+DMFT density matrix. We assume that the two-particle vertex $\Gamma$ is the same in the nematic state as in the paramagnetic state. The detailed method of computing the dynamic spin susceptibility is documented in Ref. \cite{Yin14} and was shown to be able to compute accurately the spin dynamics of many iron pnictide superconductors. To simulate the 
nematic state, 50 meV and 80 meV on-site energy splitting of the \dxz and \dyz orbitals were added to the DFT+DMFT converged solutions of the paramagnetic state for NaFeAs and BaFe$_2$As$_2$, respectively, consistent with ARPES measurements at the (1,0) and (0,1) electron pockets \cite{ZHYJ12,YLMK12}. The experimental crystal structures of NaFeAs \cite{Parker2009} and BaFe$_2$As$_2$ \cite{Rotter2008} were used in the calculations.

Figures 6(a), 6(c), and 6(e) show the calculated spin wave dispersions of 
NaFeAs from the \dxz, \dyz, and \dxy orbital channels, respectively. 
Figures 6(b), 6(d), and 6(f) show similar calculation for BaFe$_2$As$_2$. The data points overlayed
in Figs. 6(c) and 6(d) are the spin wave dispersion relations of NaFeAs and BaFe$_2$As$_2$, respectively, determined from neutron scattering experiments \cite{Lu18}. By comparing Fig. 6 with Fig. 3 for NaFeAs 
and those of detwinned
BaFe$_2$As$_2$ \cite{Lu18}, we conclude that spin waves below $E_\text{crossover} \approx 100$ meV along the 
$\mathbf{Q}_\text{strong}$ direction arise predominantly from the \dyz channel.  For $E \gtrsim 110$ meV, the neutron data shows most spectral weight at the ${\bf Q}_{xy}=(1,1)$ position [Fig. 3(f)], consistent with the calculations in the \dxy channel but not the \dyz channel.  This is also the case for spin excitations at $E=140\pm 10$ meV [Fig. 5(f)]. 
The spin excitations centered at ${\bf Q}_{xy}=(1,1)$ remain partially anisotropic at least as high as 150 meV, 
probably arising from the applied pressure-induced lattice anisotropy. There are also experimental \cite{ZHYJ12,MYi2019} and theoretical \cite{CKAF16}
evidence that the \dxy-dominated bands may participate in the broken $C_4$ rotational symmetry.

To quantitatively compare the observed spin wave anisotropy in NaFeAs with DFT+DMFT calculations, 
 we show the local dynamic susceptibility $\chi^{\prime\prime}(E)$ in Figure 7(a) as a function of energy, integrated over the green
[$\mathbf{Q}_\text{strong}=(\pm 1,0)$] and red [$\mathbf{Q}_\text{weak}=(0,\pm1)$] diamond-shaped zones shown in Fig. 1(e).
Each diamond has the area of one magnetic unit cell.
The intensity in each region is converted into absolute units of magnetic susceptibility by fitting and subtracting the background and correcting for the magnetic form factor of Fe$^{2+}$ and the Bose factor, as described in the supplementary information \cite{SI}.
The resulting integrated intensity from the strong and weak positions can be averaged to roughly match previously published data \cite{Zhang14,Carr16}.
The green, red, and blue dashed lines represent the DFT+DMFT results from  
the \dyz, \dxz, and \dxy intra-orbital scattering.  
Experimental results from SEQUOIA and MERLIN are clearly marked.

Figure 6(b) shows the spin wave 
energy dependent anisotropy, defined as $\delta=(I_s-I_w)/(I_s+I_w)$ \cite{Song2015}, and its comparison
with $\delta$ from the DFT+DMFT calculation for different orbitals. 
While the observed spin wave anisotropies are similar for energies
below $E\approx 80$ meV for SEQUOIA and MERLIN measurements, 
they are slightly different at higher energies, due in part to different 
applied uniaxial pressures in these two experiments.
The spin wave anisotropy decreases with increasing energy, and changes from 
the $C_2$ to $C_4$ rotational symmetry around 
$E_\text{crossover} \approx 100$ meV (Figs. 3-5). While the low-energy spin
excitations change anisotropy at temperatures slightly above $T_s$, spin excitations at
energies above $E_\text{crossover}$ have approximate $C_4$ rotational symmetry that does not change across $T_s$ (Figs. 4, 5). 
These results clearly reveal orbital
selective spin waves, suggesting that low-energy spin waves are mostly controlled by the \dyz orbital scattering associated 
with the nematic order below $T_s$, while high-energy spin waves near the zone 
boundary are determined by the electronic bandwidth of the 
\dxy orbitals \cite{YZSL17}.  The electronic nematic phase below $T_s$ is 
associated with the 
splitting of the \dyz and \dxz orbital bands and spin wave anisotropy between  
$\mathbf{Q}_\text{strong}$ and $\mathbf{Q}_\text{weak}$ \cite{ZHYJ12,YLMK12}, 
although such an effect
is already present above $T_s$ due to 
applied uniaxial pressure (Fig. 4) \cite{YZSL17,HRMan18}.  
 In the paramagnetic state at a temperature well above $T_s$, spin wave anisotropy between $\mathbf{Q}_\text{strong}$ and $\mathbf{Q}_\text{weak}$ becomes weakly temperature dependent together with the vanishing splitting of 
the \dyz and \dxz orbital bands \cite{YZSL17,HRMan18}.  Therefore, the energy splitting of the \dyz and \dxz orbital bands
is directly associated with the electronic and spin nematic phase, and 
spin waves in NaFeAs are orbital selective and controlled by the electronic properties of the system.

The observation of low-energy \dyz-\dyz fluctuations is consistent with the idea that these excitations are itinerant and responsible for superconducting pairing \cite{TChen2019,WYwang2017,Tian19}, while the \dxy-\dxy fluctuations are more localized and do not observably change across the electron-doped superconducting dome \cite{Carr16}.
Since the spin fluctuation spectrum in \BFA also exhibits qualitatively similar results as \NFA, and with DFT+DMFT calculations showing high-energy scattering centered at ${\bf Q}_{xy}=(1,1)$ but extending to higher energy in \BFA, we conclude that the overall magnetic bandwidth in iron pnictides is controlled by localized magnetic superexchange interactions.
With increased pnictogen height above the iron plane
and resulting increased electron correlations on moving from BaFe$_2$As$_2$ to NaFeAs, we see a clear reduction in  
energy for spin excitations at the ${\bf Q}_{xy}=(1,1)$ position,
suggesting that spin excitations around the ${\bf Q}_{xy}=(1,1)$ position are mostly associated with 
the \dxy orbital \cite{WKRE16,ZYLL16,ZHYJ12,YLMK12}.  These results are consistent with the notion that 
increased electron correlations in NaFeAs are accompanied by reduced electronic bandwidth in the \dxy orbital \cite{YZSL17} and spin wave energy bandwidth 
(and the magnetic exchange coupling $J$) \cite{Zhang14}.

Our conclusion that the microscopic origin of magnetic scattering is orbitally selective is not in contradiction with other models previously used for the iron pnictides. 
Although spin excitations in detwinned NaFeAs can be much better 
understood by orbital-selective 
scattering processes, one should also be able to parametrize these excitations with
a Heisenberg Hamiltonian which has no information concerning the orbital nature of the excitations \cite{Zhang14}. Since superconductivity in iron pinctides arises from 
electron/hole-doped antiferromagnets, one can use the 
$t$-$J$ model to estimate high temperature superconductivity states regardless of the microscopic origin of spin excitations \cite{Karchev98}. In the notion of spin
fluctuation driven superconductivity \cite{scalapinormp}, the superconducting condensation energy of the system should be dominated by the change in magnetic 
exchange energy between the normal (N) and
superconducting (S) phase at zero temperature $\Delta E_{ex}(T=0)$. Within the 
$t$-$J$ model, this means that 
$\Delta E_{ex}(T)=2J[\left\langle {\bf S}_{i+x}\cdot {\bf S}_{i}\right\rangle_{\rm N}-\left\langle {\bf S}_{i+x}\cdot {\bf S}_{i}\right\rangle_{\rm S}]$, 
where $J$ is the nearest neighbor magnetic exchange coupling and $\left\langle {\bf S}_{i+x}\cdot {\bf S}_{i}\right\rangle$ is the magnetic scattering in
absolute units at temperature $T$ \cite{MWang2013}.  In this picture, the quasiparticle excitations across the nested hole-electron Fermi surfaces in the nematic phase arise mostly
from the \dyz-\dyz intra-orbital excitations, and give rise to the $C_2$ neutron spin resonance seen in different families of iron-based superconductors \cite{TChen2019,WYwang2017,Tian19}.  The increased electron-electron correlations from BaFe$_2$As$_2$ to NaFeAs, reflected in reduced bandwidth of the \dxy orbital and high-energy spin excitations \cite{Zhang14}, correspond to the reduced magnetic exchange coupling $J$.  Therefore, localized electrons in the \dxy orbital and associated high energy spin excitations with $C_4$ symmetry are also important for superconductivity of iron-based materials.

\section{Conclusions}

In summary, we have used inelastic neutron scattering to study spin waves of detwinned crystals of NaFeAs. By comparing 
the energy and temperature dependence of the spin wave anisotropy with DFT+DMFT calculations, we conclude that spin waves in iron pnictides are orbital selective.   While the 
\dxy-orbital scattering processes control the bandwidth of spin fluctuations in the iron pnictides, the low energy spin fluctuations may arise through Fermi surface nesting of itinerant electrons with 
\dyz orbital character below $T_s$, and coupled to the nematic phase transition. Superconductivity in nematic ordered iron pnictides may therefore be controlled by Fermi surface nesting of itinerant electrons with
\dyz orbital character while high energy spin excitations associated with the \dxy orbital and electron correlations are also important.

\section{Acknowledgments}

The neutron scattering work at Rice is supported by the
U.S. Department of Energy (DOE), Basic Energy Sciences
(BES), under Contract No. DE-SC0012311 (P.D.)  The materials synthesis efforts at Rice is supported by the Robert A. Welch Foundation Grant Nos. C-1839 (P.D.). The research at ORNL was sponsored by the Scientific User Facilities Division, Office of BES, U.S. DOE. ZPY was supported by the NSFC (Grant No. 11674030), the Fundamental Research Funds for the Central Universities (Grant No. 310421113), the National Key Research and Development Program of China grant 2016YFA0302300. The calculations used high performance computing clusters at BNU in Zhuhai and the National Supercomputer Center in Guangzhou. MY is supported by the the Robert A. Welch Foundation Grant No. C-2024 as well as the Alfred P. Sloan Foundation.


\begin{thebibliography}{}

\bibitem{scalapinormp} D. J. Scalapino, Rev. Mod. Phys. {\bf 84}, 1383 (2012).
\bibitem{dairmp} P. C. Dai, Rev. Mod. Phys. {\bf 87}, 855 (2015).
\bibitem{palee} P. A. Lee, N. Nagaosa, and X. G. Wen, Rev. Mod. Phys. {\bf 78}, 17 (2006).
\bibitem{scalapino} D. J. Scalapino, Physics Reports {\bf 250}, 329 (1995).
\bibitem{stewart} G. R. Stewart, Rev. Mod. Phys. {\bf 83}, 1589-1652 (2011).
\bibitem{hirschfeld} P. J. Hirschfeld, M. M. Korshunov, and I. I. Mazin,
Rep. Prog. Phys. {\bf 74}, 124508 (2011).
\bibitem{QSi} Q. Si, R. Yu, and E. Abrahams, Nat. Rev. Mater. {\bf 1}, 16017 (2016).
\bibitem{YZSL17} M. Yi, Y. Zhang, Z.-X. Shen, and D. Lu. Npj Quantum Materials {\bf 2}, 57 (2017).
\bibitem{Yin11} Z. P. Yin, K, Haule, and G. Kotliar, Nat. Mater. {\bf 10}, 932 (2011).
\bibitem{Fernandes} R. M. Fernandes, A. V. Chubukov, and J. Schmalian, Nat. Phys. {\bf 10}, 97 (2014).
\bibitem{WKRE16} M. D. Watson, T. K. Kim, L. C. Rhodes, M. Eschrig, M. Hoesch, A. A. Haghighirad, and A. I. Coldea. Phys. Rev. B {\bf 94}, 201107 (2016).
\bibitem{FMTC16} L. Fanfarillo, J. Mansart, P. Toulemonde, H. Cercellier, P. Le F$\rm \grave{e}$vre, F. Bertran, B. Valenzuela, L. Benfatto, and V. Brouet. Phys. Rev. B {\bf 94}, 155138 (2016).
\bibitem{Hsu} Fong-Chi Hsu, Jiu-Yong Luo, Kuo-Wei Yeh, Ta-Kun Chen, Tzu-Wen Huang, Phillip M. Wu, Yong-Chi Lee, Yi-Lin Huang, Yan-Yi Chu, Der-Chung Yan, and Maw-Kuen Wu,  
Proc. Natl Acad. Sci. USA {\bf 105}, 14262 (2008).
\bibitem{McQueen09} T. M. McQueen, A. J. Williams, P. W. Stephens, J. Tao, Y. Zhu, V. Ksenofontov, F. Casper, C. Felser, and R. J. Cava, Phys. Rev. Lett. {\bf 103}, 057002 (2009).
\bibitem{bohmer} A. E. B\"{o}hmer, and A. Kreisel, J. Phys.: Condens. Matter {\bf 30}, 023001 (2018).

\bibitem{YLZY15} M. Yi, Z.-K. Liu, Y. Zhang, R. Yu, J.-X. Zhu, J. J. Lee, R. G. Moore, F. T. Schmitt, W. Li, S. C. Riggs, J.-H. Chu, B. Lv, J. Hu, M. Hashimoto, S.-K. Mo, Z. Hussain, Z. Q. Mao, C. W. Chu, I. R. Fisher, Q. Si, Z.-X. Shen, and D. H. Lu, Nature Communications {\bf 6}, 7777 (2015).

\bibitem{HMiao2016} H. Miao, Z. P. Yin, S. F. Wu, J. M. Li, J. Ma, B.-Q. Lv, X. P. Wang, T. Qian, P. Richard, L.-Y. Xing, X.-C. Wang, C. Q. Jin, K. Haule, G. Kotliar, and H. Ding, 
Phys. Rev. B {\bf 94}, 201109(R) (2016).

\bibitem{Yin2012} Z. P. Yin, K. Haule, and G. Kotliar, 
Phys. Rev. B {\bf 86}, 195141 (2012).

\bibitem{Sprau} P. O. Sprau, A. Kostin, A. Kreisel, A. E. B\"{o}hmer, V. Taufour,
P. C. Canfield, S. Mukherjee, P. J. Hirschfeld, B. M. Andersen,
and J. C. S$\rm \acute{e}$amus Davis, Science {\bf 357}, 75 (2017).

\bibitem{ChKF16} A. V. Chubukov, M. Khodas, and R. M. Fernandes. Phys. Rev. X {\bf 6}, 041045 (2016).

\bibitem{TChen2019} Tong Chen, Youzhe Chen, Andreas Kreisel, Xingye Lu, Astrid Schneidewind, Yiming Qiu, J. T. Park, Toby G. Perring, J Ross Stewart, Huibo Cao, Rui Zhang, Yu Li, Yan Rong, Yuan Wei, Brian M. Andersen, P. J. Hirschfeld, Collin Broholm, and Pengcheng Dai, Nat. Mater. {\bf 18}, 709 (2019). 

\bibitem{WYwang2017} Weiyi Wang, J. T. Park, Rong Yu, Yu Li, Yu Song, Zongyuan Zhang, Alexandre Ivanov, Jiri Kulda, and Pengcheng Dai, Phys. Rev. B {\bf 95}, 094519 (2017).

\bibitem{Tian19} Long Tian, Panpan Liu, Zhuang Xu, Yu Li, Zhilun Lu, H. C. Walker, U. Stuhr, Guotai Tan, Xingye Lu, and Pengcheng Dai, Phys. Rev. B {\bf 100}, 134509 (2019). 

\bibitem{Eschrig} M. Eschrig, Adv. Phys. {\bf 55}, 47-183 (2006).

\bibitem{Maier} T. A. Maier and D. J. Scalapino, Phys. Rev. B {\bf 78}, 020514(R) (2008). 

\bibitem{RZhang} Rui Zhang, Weiyi Wang, Thomas A. Maier, Meng Wang, Matthew B. Stone, Songxue Chi, Barry Winn, and Pengcheng Dai, Phys. Rev. B {\bf 98}, 060502(R) (2018).

\bibitem{Inosov2010} D. S. Inosov, J. T. Park, P. Bourges, D. L. Sun, Y. Sidis, A. Schneidewind, K. Hradil, D. Haug, C. T. Lin, B. Keimer, and V. Hinkov, Nat. Phys. {\bf 6}, 178 (2010).

\bibitem{CHLee2013} C. H. Lee, P. Steffens, N. Qureshi, M. Nakajima, K. Kihou, A. Iyo, H. Eisaki, and M. Braden, Phys. Rev. Lett. {\bf 111}, 167002 (2013).

\bibitem{Fwaber2019} F. Wa$\rm \beta$er, J. T. Park, S. Aswartham, S. Wurmehl, Y. Sidis, P. Steffens, K. Schmalzl, B. B$\rm \ddot{u}$chner, and M. Braden, npj Quantum Materials {\bf 4}, 59 (2019).


\bibitem{Zhang14} C. Zhang, L. W. Harriger, Z. Yin, W. Lv, M. Wang, G. Tan, Y. Song, D. L. Abernathy, W. Tian, T. Egami, K. Haule, G. Kotliar, and P. Dai, Phys. Rev. Lett. {\bf 112}, 217202 (2014).

\bibitem{JZhao09} J. Zhao, D. T. Adroja, D.-X. Yao, R. Bewley, S. Li, X. F.
Wang, G. Wu, X. H. Chen, J. Hu, and P. C. Dai, Nat. Phys. {\bf 5}, 555 (2009).

\bibitem{leland11} L. W. Harriger, H. Q. Luo, M. S. Liu, C. Frost, J. P. Hu, M. R. Norman, and Pengcheng Dai, Phys. Rev. B {\bf 84}, 054544 (2011).

\bibitem{Ewings11} R. A. Ewings, T. G. Perring, J. Gillett, S. D. Das, S. E.
Sebastian, A. E. Taylor, T. Guidi, and A. T. Boothroyd,
Phys. Rev. B {\bf 83}, 214519 (2011).


\bibitem{Yin14} Z. P. Yin, K. Haule, G. Kotliar, Nat. Phys. {\bf 10}, 845 (2014).

\bibitem{Lu18} X. Lu, D. D. Scherer, D. W. Tam, W. Zhang, R. Zhang, H. Luo, L. W. Harriger, H. C. Walker, D. T. Adroja, B. M. Andersen, and P. Dai, Phys. Rev. Lett. {\bf 121}, 067002 (2018).

\bibitem{qhuang} Q. Huang, Y. Qiu, Wei Bao, M. A. Green, J. W. Lynn, Y. C. Gasparovic, T. Wu, G. Wu, X. H. Chen,  Phys. Rev. Lett. {\bf 101}, 257003 (2008).

\bibitem{kim2011} M. G. Kim, R. M. Fernandes, A. Kreyssig, J. W. Kim, A. Thaler, S. L. Bud'ko, P. C. Canfield, R. J. McQueeney, J. Schmalian, A. I. Goldman,  Phys. Rev. B {\bf 83}, 134522 (2011).

\bibitem{SLi09} Shiliang Li, Clarina de la Cruz, Q. Huang, G. F. Chen, T.-L. Xia, J. L. Luo, N. L. Wang, and Pengcheng Dai, Phys. Rev. B {\bf 80}, 020504(R) (2009). 

\bibitem{Tan16} Guotai Tan, Yu Song, Chenglin Zhang, Lifang Lin, Zhuang Xu, Tingting Hou, Wei Tian, Huibo Cao, Shiliang Li, Shiping Feng, and Pengcheng Dai, Phys. Rev. B {\bf 94}, 014509 (2016).

\bibitem{SQUOIA} G. E. Granroth, A. I. Kolesnikov, T. E. Sherline, J. P. Clancy, K. A. Ross, J. P. C. Ruff, B. D. Gaulin, S. E. Nagler, Journal of Physics: Conference Series {\bf 251}, 12058 (2010).

\bibitem{MERLIN} R. I. Bewley, T. Guidi, and S. Bennington, Notiziario Neutroni e Luce 
di Sincrotrone {\bf 14}, 22 (2009).

\bibitem{SI} See supplementary information for additional data and analysis, which include 
Refs. \cite{Tam2019b,QWang2016}. 

\bibitem{Dhital2012} C. Dhital, Z. Yamani, W. Tian, J. Zeretsky, A. S. Sefat, Z. Wang, R. J. Birgeneau, and S. D. Wilson, Phys. Rev. Lett. {\bf 108}, 087001 (2012).
\bibitem{Dhital2014} C. Dhital, T. Hogan, Z. Yamani, R. J. Birgeneau, W. Tian, M. Matsuda, A. S. Sefat, Z. Wang, and S. D. Wilson, Phys. Rev. B 89, 214404 (2014).

\bibitem{YSong2013} Y. Song, S. V. Carr, X. Y. Lu, C. L. Zhang, Z. C. Sims, N. F. Luttrell, S. X. Chi, Y. Zhao, J. W. Lynn, and P. C. Dai, Phys. Rev. B {\bf 87}, 184511 (2013).

\bibitem{Tam2017} David W. Tam, Yu Song, Haoran Man, Sky C. Cheung, Zhiping Yin, Xingye Lu, Weiyi Wang, Benjamin A. Frandsen, Lian Liu, Zizhou Gong, Takashi U. Ito, Yipeng Cai, Murray N. Wilson, Shengli Guo, Keisuke Koshiishi, Wei Tian, Bassam Hitti, Alexandre Ivanov, Yang Zhao, Jeffrey W. Lynn, Graeme M. Luke, Tom Berlijn, Thomas A. Maier, Yasutomo J. Uemura, and Pengcheng Dai, Phys. Rev. B {\bf 95}, 060505(R) (2017). 

\bibitem{Lu14} X. Lu, J. T. Park, R. Zhang, H. Luo, A. H. Nevidomskyy, Q. Si, and P. C. Dai, Science {\bf 345}, 657 (2014).

\bibitem{ZYLL16} Y. Zhang, M. Yi, Z.-K. Liu, W. Li, J. J. Lee, R. G. Moore, M. Hashimoto, M. Nakajima, H. Eisaki, S.-K. Mo, Z. Hussain, T. P. Devereaux, Z.-X. Shen, and D. H. Lu. Phys. Rev. B {\bf 94}, 115153 (2016).

\bibitem{ZHYJ12} Y. Zhang, C. He, Z. R. Ye, J. Jiang, F. Chen, M. Xu, Q. Q. Ge, B. P. Xie, J. Wei, M. Aeschlimann, X. Y. Cui, M. Shi, J. P. Hu, and D. L. Feng. Phys. Rev. B {\bf 85}, 085121 (2012).

\bibitem{YLMK12} M. Yi, D. H. Lu, R. G. Moore, K. Kihou, C.-H. Lee, A. Iyo, H. Eisaki, T. Yoshida, A. Fujimori, and Z.-X. Shen. New J. Phys. {\bf 14}, 073019 (2012).

\bibitem{JHZhang2010} Junhua Zhang, Rastko Sknepnek, and J$\rm \ddot{o}$rg Schmalian, Phys. Rev. B {\bf 82}, 134527 (2010).

\bibitem{Johannes06} M. D. Johannes, I. I. Mazin, and C. A. Howells, Phys. Rev. B {\bf 73}, 205102 (2006). 

\bibitem{LYSW18} Y. Li, Z. Yamani, Y. Song, W. Wang, C. Zhang, D. W. Tam, T. Chen, D. Hu, Z. Xu, S. Chi, K. Xia, L. Zhang, S. Cui, W. Guo, Z. Fang, Y. Liu, and P. Dai. Phys. Rev. X {\bf 8}, 021056 (2018).

\bibitem{Chu10} Jiun-Haw Chu, James G. Analytis, Kristiaan De Greve, Peter L. McMahon,
Zahirul Islam, Yoshihisa Yamamoto, and Ian R. Fisher, Science {\bf 329}, 824 (2010). 

\bibitem{HRMan18} H. R. Man, R. Zhang, J. T. Park, Xingye Lu, J. Kulda, A. Ivanov, and P. C. Dai, Phys. Rev. B {\bf 97}, 060507(R) (2018).

\bibitem{DWTam2019} D. W. Tam, W. Y. Wang, L. Zhang, Y. Song, R. Zhang, S. V. Carr, H. C. Walker, T. G. Perring, D. T. Adroja, and P. C. Dai, Phys. Rev. B {\bf 99}, 134519 (2019). 

\bibitem{XYLu2016} X.Y. Lu, K.-F. Tseng, T. Keller, 
W. L. Zhang, D. Hu, Y. Song, H. R. Man, J. T. Park, H. Q. Luo, S. L. Li, A. H. Nevidomskyy, and P. C. Dai, Phys. Rev. B {\bf 93}, 134519 (2016).  

\bibitem{Kotliar2006} G. Kotliar, S. Y. Savrasov, K. Haule, V. S. Oudovenko, O. Parcollet, and C. A. Marianetti, Rev. Mod. Phys. {\bf 78}, 865 (2006).

\bibitem{Blaha2001} P. Blaha, K. Schwarz, G. Madsen, D. Kvasnicka, and J. Luitz, WIEN2K, An Augmented Plane Wave+Local Orbitals Program for Calculating Crystal Properties (Karlheinz Schwarz, Techn. Universit$\rm \ddot{a}$t Wien, Austria, 2001).

\bibitem{Perdew1996} J. P. Perdew, K. Burke, and M. Ernzerhof, Phys. Rev. Lett. {\bf 77}, 3865 (1996).

\bibitem{Haule2010} K. Haule, C.-H. Yee, K. Kim Phys. Rev. B {\bf 81}, 195107 (2010).

\bibitem{Haule2007} K. Haule, Phys. Rev. B {\bf 75}, 155113 (2007).

\bibitem{Werner2006} P. Werner, A. Comanac, L. de' Medici, M. Troyer, and A. J. Millis, Phys. 
Rev. Lett. {\bf 97}, 076405 (2006).

\bibitem{Parker2009} D. R. Parker, M. J. Pitcher, P. J. Baker, I. Franke, T. Lancaster, S. J. Blundell, and S. J. Clarke, Chem. Commun., 2189-2191 (2009).  

\bibitem{Rotter2008} M. Rotter, M. Tegel, D. Johrendt, I. Schellenberg, W. Hermes, and R. P$\rm \ddot{o}$ttgen, 
Phys. Rev. B {\bf 78}, 020503(R) (2008).

\bibitem{MYi2019} M. Yi, H, Pfau, Y, Zhang, Y, He, H, Wu, T. Chen, Z.R. Ye, M. Hashimoto, R. Yu, Q. Si, D.-H. Lee, Pengcheng Dai, Z.-X. Shen, D. H. Lu, and R. J. Birgenau, Phys. Rev. X {\bf 9}, 041049 (2019). 

\bibitem{CKAF16} M. H. Christensen, J. Kang, B. M. Andersen, and R. M. Fernandes. Phys. Rev. B {\bf 93}, 085136 (2016).

\bibitem{Carr16} S. V. Carr, C. Zhang, Y. Song, G. Tan, Y. Li, D. L. Abernathy, M. B. Stone, G. E. Granroth, T. G. Perring, and P. Dai, Phys. Rev. B {\bf 93}, 214506 (2016).

\bibitem{Song2015} Yu Song, Xingye Lu, D. L. Abernathy, David W. Tam, J. L. Niedziela, Wei Tian, Huiqian Luo, Qimiao Si, and Pengcheng Dai, Phys. Rev. B {\bf 92}, 180504(R) (2015). 

\bibitem{Karchev98} Naoum Karchev, Phys. Rev. B {\bf 57}, 10913 (1998). 

\bibitem{MWang2013} M. Wang, C. Zhang, X. Lu, G. T. Tan, H. Luo, Y. Song, M. Y. Wang, X. Zhang, E.A. Goremychkin, T. G. Perring, T. A. Maier, Z. P. Yin, K. Haule, G. Kotliar, P. Dai, Nat. Comm. {\bf 4}, 2874 (2013).

\bibitem{Tam2019b} D. W. Tam, H.-H. Lai, J. Hu, X. Lu, H. C. Walker, D. L. Abernathy, J. L. Niedziela, T. Weber, M. Enderle, Y. Su, Z. Q.
Mao, Q. Si, and P. Dai, Phys. Rev. B {\bf 100}, 054405 (2019).

\bibitem{QWang2016} Q. Wang, Y. Shen, B. Pan, X. Zhang, K. Ikeuchi, K. Iida, A. D. Christianson, H. C. Walker, D. T. Adroja, M. Abdel-Hafiez,
X. Chen, D. A. Chareev, A. N. Vasiliev, and J. Zhao, Nat. Comm. {\bf 7}, 12182 (2016).

\end{thebibliography}
\end{document}